\documentclass[prd,aps,floats,showpacs,twocolumn,twoside,preprintnumbers,
superscriptaddress,floatfix,11pt]{revtex4}

\usepackage{slashed}
\usepackage{graphicx}
\usepackage{pstricks}

\newcommand{\gsim}{\stackrel{\scriptstyle >}{\phantom{}_{\sim}}}

\newcommand{\MSun}{{\rm M}_\odot}

\begin{document}

\title{
Implications of 
the measurement of pulsars with two solar masses
for quark matter in compact stars and HIC. 
A NJL model case study.
}

\author{T. Kl\"ahn}
\email{thomas.klaehn@ift.uni.wroc.pl}
\affiliation{Instytut Fizyki Teoretycznej, Uniwersytet Wroc{\l}awski, 
PL-50-204 Wroc{\l}aw, Poland}

\author{R. {\L}astowiecki}
\email{lastowiecki@ift.uni.wroc.pl}
\affiliation{Instytut Fizyki Teoretycznej, Uniwersytet Wroc{\l}awski, 
PL-50-204 Wroc{\l}aw, Poland}

\author{D. Blaschke}
\email{blaschke@ift.uni.wroc.pl}
\affiliation{Instytut Fizyki Teoretycznej, Uniwersytet Wroc{\l}awski, 
PL-50-204 Wroc{\l}aw, Poland}
\affiliation{Bogoliubov Laboratory for Theoretical Physics, JINR  
Dubna, RU-141980 Dubna, Russia}
\affiliation{Fakult\"at f\"ur Physik, Universit\"at Bielefeld, 
D-33615 Bielefeld, Germany}

\date{\today}
\begin{abstract}
The precise measurement of the high 
masses of the pulsars 
PSR J1614-2230 ($M_{1614}=1.97 \pm 0.04~M_\odot$) and 
PSR J0348-0432 ($M_{0348}=2.01 \pm 0.04~M_\odot$) 
provides an important constraint for the equation of state of cold, 
dense matter and is suited to give interesting insights regarding 
the nature and existence of the possible phase transition to
deconfined quark matter {in the cores of neutron stars}.
We
analyze the stability and composition of 
compact star sequences for a class of hybrid 
nuclear -- quark-matter equations of state.
The quark matter phase is described in the framework of a standard
color superconducting 3-flavor Nambu--Jona-Lasinio model
and the hadronic phase is
given by the Dirac-\-Brueckner-\-Hartree-\-Fock equation of state for the 
Bonn-A potential.
The phase transition is obtained by a Maxwell construction.
Within this model setup we aim to
constrain otherwise not strictly fixed parameters of the NJL model, namely
the coupling strengths in the vector meson and diquark interaction channels.
We perform this investigation for two different parameterizations
characterized by a different scalar coupling constant.
The analysis of flow data obtained in heavy-ion 
collisions resulted in a further constraint which we account for in our
discussion.
Massive hybrid stars with extended quark matter cores can be obtained
in accordance with all of the considered constraints.
\end{abstract}

\maketitle

\section{Introduction}
Neutron stars (NS) are considered to be cosmic laboratories for dense matter
\cite{Weber:1999qn,Blaschke:2009}.
Of particular interest is the fact that the region of high densities and very 
low temperatures where NS are located in the QCD phase diagram is not 
accessible for terrestrial experiments or lattice simulations of QCD.
The physics of NS is studied intensively in order to derive constraints for 
theories of high density physics
aiming to complement the insights obtained from heavy-ion collisions.
In the effort to understand the complex physics of neutron stars a large variety of observables
has been studied which provide valuable constraints on the equation of state (EoS). For recent reviews see, e.g.,  
\cite{Miller:2002qw,Bhattacharyya:2010fn,Blaschke:2009,Truemper:2011,Lattimer:2012nd}.
A crucial observable in this investigation is the maximum attainable mass 
of a NS which is directly connected to the EoS.
The precise knowledge of the highest NS mass
puts significant constraints
on the stiffness of the EoS and can 
rule out entire classes of EoS models\cite{Lattimer:2010uk}.
More advanced approaches aim to process simultaneously 
spectra, luminosities and distances
of as many as possible NS in order
to extract masses and radii simultaneously. These information
can be further evaluated to determine a
most probable underlying EoS 
within a Bayesian framework 
\cite{Steiner:2010fz,Steiner:2012xt}.
While this method promises a detailed reconstruction of the EoS it relies on 
the availability of high quality, indisputable data suitable for the
extraction of mass and radius constraints.
Efforts are being made in the astronomers community to achieve 
this 
goal \cite{Arzoumanian:2009qn}.

The recent measurement of 
two massive
neutron stars with
$M_{1614}=1.97\pm 0.04~\MSun$ 
(for PSR J1614-2230 \cite{Demorest:2010bx}) and 
$M_{0348}=2.01\pm 0.04~\MSun$ (for PSR J0348-0432 \cite{Antoniadis:2013pzd})
revived the discussion about possible implications of a NS with about two 
solar masses
for the equation of state of cold and dense matter \cite{Ozel:2010bz,Lattimer:2010uk}.
These well measured masses greatly exceed the previously 
highest well known NS mass of $1.667 \pm 0.021~M_\odot$ 
for PSR J1903+0327 \cite{Champion:2008,Freire:2010tf}.
Because of 
the narrow error bands already this object provided a strong constraint 
on the stiffness of the equation of state.
For example, it was suggested that this measurement puts 
a considerable strain on the possibility of the existence 
of quark matter in the cores of dense stellar 
objects \cite{Lattimer:2010uk}.
A discussion  similar to todays regarding PSR J1614-2230
followed when a pulsar mass of about $2M_{\odot}$ had been reported for
J0751+1807 \cite{Nice:2005fi}, see \cite{Klahn:2006iw} and references therein.
Although this value had to be corrected afterwards,
all conclusions drawn from the mere fact, that such 
heavy NS 
exist still stand. 
Contrary to other claims (e.g., Ref.~\cite{Ozel:2006bv})
we emphasize that observations of NS with masses of $2~M_{\odot}$ 
and beyond do not exclude the existence of hybrid NS with a quark matter 
core \cite{Alford:2006vz}. 
Partially, we use the present work to reemphasize this known fact 
within the framework of an Nambu--Jona-Lasinio (NJL) model 
in a similar but more detailed analysis as we
performed it in previous work\cite{Klahn:2006iw}.

The appearance of new degrees of freedom, the most prominent being hyperons 
and quarks, 
entails a softening of the EoS and
in general reduces the maximum NS mass in comparison to the
underlying pure nuclear matter EoS.
This softening, however, is not necessarily strong enough to 
conclude
that NS masses as high as $2~M_\odot$ preclude the existence of exotic matter in the NS core.
This has been confirmed in a number of studies concerning both, 
the occurrence of hyperons (see, e.g., 
\cite{Weissenborn:2011kb,Bednarek:2011gd,Sulaksono:2012ny,Katayama:2012ge,Lastowiecki:2011hh})  
and the transition to 
quark matter described
within various different model approaches, e.g., 
\cite{Weissenborn:2011qu,Klahn:2011fb,Sagert:2011kf,Bonanno:2011ch,Lenzi:2012xz,Chamel:2012ea,Franzon:2012in,Kim:2012pr,Mallick:2012wb,Masuda:2012kf}.
The discussion of the limits on the stiffness of the high-density EoS is also 
performed in the context of heavy-ion collision experiments which provide 
further constraints  on the EOS stiffness, see \cite{Sagert:2011kf,Klahn:2011}.

In the present paper we systematically scan a part of the
NJL model parameter space in order to locate those parameter regions
which result in QM cores for massive NS.
We fully scan the region of vector- and diquark couplings
which result in stable NS configurations with QM cores.
Additionally, we apply two different parameterizations regarding
the scalar coupling, which both reproduce the
pion and kaon mass as well 
as
the pion decay constant and light quark mass in vacuum. 
The differences between them result
from slightly different choices for the scalar coupling
constant which are compensated by a different three-momentum cutoff $\Lambda$. Details concerning the parameterization scheme are found in 
\cite{Grigorian:2006qe}.
Additionally, we investigate 
how these hybrid EoS which we find to favor massive NS with QM core
agree with the flow constraint \cite{Danielewicz:2002pu}.

In order to keep this study sufficiently transparent
we have chosen to vary the free parameters of the EoS model 
only in the quark sector
and to  apply the 
{\it ab-initio} Dirac-\-Brueckner-\-Hartree-\-Fock (DBHF)
EoS using the Bonn-A nucleon-nucleon potential \cite{Fuchs:2003zn}
as the only {\it nuclear} matter EoS we investigate. 
The latter  well describes 
the saturation properties of symmetric nuclear matter,
provides 
a sufficiently high maximum NS mass of $2.4~M_\odot$, and is in 
agreement with the flow constraint 
up to 3.5 times saturation density \cite{Klahn:2006ir}.
The phase transition between nuclear and quark matter is modelled
in terms of a Maxwell construction.

This work is structured in the following way: 
Sect.~II discusses 
the NJL model and
a variety of coupling channels one {\it could} account for.
While we do not consider all of these channels we find it instructive
to discuss the variety NJL-type models offer.
Later in the section we focus on the model as it is used for this
study, namely accounting for the scalar-, vector-, and diquark- 
interaction channel in the mean field approximation.
In Sect. III we discuss the obtained compact star sequences and the
agreement of the corresponding EoS parameterizations with flow data
in symmetric matter.
Section IV discusses our conclusions
of this study.

\section{ Dense hybrid star matter }
\subsection{NJL-type quark matter models}
The NJL model 
has originally been introduced
by Nambu and Jona--Lasinio 
as a field theoretical model to understand 
the origin of the mass of nucleons
as a selfenergy in a theory with four-fermion interactions
in analogy to the occurrence of an energy 
gap in the theory of superconductivity \cite{Nambu:1961tp}. 
Nowadays this model is widely appreciated as 
a useful tool to model the thermodynamics of 
deconfined quark matter, 
see \cite{Buballa:2003qv} for a comprehensive review
with particular emphasis on the high density aspects.
Several reasons contribute to this fact. 
First, it describes already in a very simplified form, 
which takes only the attractive scalar interaction term into account, 
one of the most prominent key features of QCD, 
the dynamical breaking of chiral symmetry.
This is a clear distinction to the thermodynamic
bag model which still is widely applied as it is
easy to use for explorative purposes.
A further advantage of NJL-type models is the
availability of a wide number of different
interaction channels which would result, 
e.g., from a global color model of QCD \cite{Tandy:1997qf}
after Fierz rearrangement of the current-current type interaction
\cite{Cahill:1988zi,Ebert:1994mf}.
Even though we strongly benefit from the NJL models 'simplicity'
we point out, that it cannot be claimed to be equivalent to QCD. 
It reproduces some of the symmetries of the full theory, but not all.
Among the missing features we mention the local color gauge symmetry. 
Further, interactions are considered on the level of one-gluon exchange.
Hence, higher order non-perturbative self-interactions are ignored
\cite{Klevansky:1992qe}.

In this section we discuss the choices
we made for the interaction part to define 
what we consider as a standard NJL model for
applications to compact star physics.
We also discuss aspects of NJL-type models which we denote as
extensions to the standard NJL model.

Of importance in the low temperature domain at
high densities is the formation of diquark condensates signalling
color superconducting properties of the system.
Solutions of the self-consistent 
meanfield equations for the quark masses and 
diquark gaps within the 3-flavor NJL model have first been presented in
\cite{Blaschke:2005uj,Ruester:2005jc,Abuki:2005ms}.
Shortly after, the before neglected vector channel interaction
has been included \cite{Klahn:2006iw}. 
This resulted in stiff QM EoS which describe
hybrid NS with QM cores 
in full agreement with the observation
of even the most massive NS - given the
nuclear matter EoS is sufficiently stiff \cite{Klahn:2006iw}.

The NJL Lagrangian for a quark matter model 
can be obtained from a global color model of QCD which ascribes all 
nonperturbative low-energy QCD aspects to the coupling of quark currents 
via a model gluon propagator which in the limit of heavy gluon exchange 
reduce to a local coupling.
Technically this is followed by a Fierz-transformation 
which results in a number of different interaction channels.
The Fierz-transformation gives an explicit ratio
between the coupling constants of the different channels,
which, if applied strictly, usually do not result in
the very best results when it comes to describing NS observables.
As it shows, already rather small variations of the
couplings do significantly change the outcome of this kind of studies.
Performing these variations seems completely legitimate to
us, considering the before stated fact, that an NJL-model
due to the underlying simplifications can at best be understood
as an effective model, which consequently should be interpreted
flexible enough to describe reality if adjusted properly.
The interplay of all possible interaction channels
so far has never been fully studied.
We conclude at the current stage of
research that one should 
systematically investigate the influence of every single one.
Further terms, which do not directly follow from the sketched
approach but can result from the full theory,
as for instance the Polyakov-loop term describing the gluon sector, 
should be carefully added.

As a standard NJL-type model Lagrangian for dense quark matter studies we define
\begin{equation}
{\cal L} = {\cal L}_0 + {\cal L}_S + {\cal L}_{PS} + {\cal L}_V 
+ {\cal L}_{AV} + {\cal L}_D 
.
\end{equation} 
The first term in the Lagrangian is the kinetic term for free Dirac quarks,
\begin{equation}
{\cal L}_{0} = \bar{q}(-i\gamma^\mu \partial_\mu +\hat{m}+ \gamma_0\hat{\mu})q~,
\end{equation}
where $\hat{m}={\rm diag}_f(m_u,m_d,m_s)$ is the current quark mass matrix,  
and $\hat{\mu}={\rm diag}_f(\mu_u,\mu_d,\mu_s)$ the corresponding matrix for 
the quark chemical potentials.
Instead of accounting explicitly for the
 minimal coupling to the non-Abelian, self-interacting gluon 
fields prominent
in the original QCD Lagrangian, we understand here that the
gluon degrees of freedom are ``integrated out'' leaving instead an interaction
Lagrangian consisting of combinations of quark bilinears with
nonperturbatively strong couplings.
The standard choice for these current-current interaction terms is guided by
the observed meson spectrum and by chiral symmetry.
The most important of these are the scalar and pseudo-scalar interaction 
channels,
\begin{eqnarray}
 {\cal L}_S &=& G_S\sum_{a=0}^{8}(\bar{q}\tau_a q)^2,\\
 {\cal L}_{PS} &=& G_{PS}\sum_{a=0}^{8}(\bar{q}\gamma_5\tau_a q)^2,
\end{eqnarray}
with the Gell-Mann matrices  $\tau_a$ acting in flavor space,
gives access to
the key-feature for which NJL-type
models are widely used, viz. the dynamical breaking of chiral 
symmetry. The pseudo-scalar channel does not contribute 
to the thermodynamical potential in meanfield approximation,
but as it is formally necessary in order to keep the Lagrangian
chirally invariant, we consider it being part of the model.
Consequently, as soon as local variations of the scalar contributions are
investigated, both terms should be taken into account.
Chiral symmetry then requires $G_{PS} = G_S$.

The next two terms describe the vector and pseudo-vector interaction,
again of current-current type.
As discussed earlier they are extremely important if one investigates
the question whether QM occurs in the interior of NS.
It has often been claimed that the existence of massive NS
implies a negative answer.
This statement, although wide spread and often met 
is nevertheless wrong. 
The reason for this wrong conclusion is usually
found in a particular choice for the model which
is supposed to describe QM at finite densities.
This manifests in an insufficient
stiffness of the EoS even though there is nor theoretical reason
for such a limitation.
In our model a stiffening is described by the repulsive 
vector (and pseudo-vector) interaction channel,
\begin{eqnarray}
 {\cal L}_V &=& G_V(\bar{q}i\gamma_\mu q)^2,\\
 {\cal L}_{AV} &=& G_{AV}(\bar{q}i\gamma_5\gamma_0q)^2.
\end{eqnarray}
Note, that these terms naturally appear after Fierz transformation
of the 
local heavy gluon exchange interaction model
which would then fix also the coupling strengths relative  
to the scalar channel as
$G_V=G_{AV}=\frac{1}{2}G_S$  (see, e.g., appendix A of \cite{Buballa:2003qv}).
As for the pseudo-scalar interaction the term related to the axial-vector
coupling vanishes in mean field approximation but is required
to keep the Lagrangian chirally symmetric.
In this sense we consider it to be part of the standard NJL model without
impact on the mean field thermodynamics.

The last term we consider accounts for scalar diquark correlations,
\begin{equation}
 {\cal L}_D = G_D\sum_{a,b=2,5,7}
(\bar{q}i\gamma_5\tau_a\lambda_b C \bar{q}^T)(q^TCi\gamma_5\tau_a\lambda_a q)~,
\label{diquark}
\end{equation}
where $\tau_a$ are again the Gell-Mann matrices in flavor space, but we 
also introduce their counterparts $\lambda_a$ acting in the color space.
The matrix $C$ is the charge conjugation matrix and $G_D$ is yet another 
coupling constant introduced into the model.
Note, that this term involves an interaction vertex of the $(qq)^2$ type
and that it is antisymmetric w.r.t. quark exchange thus fulfilling the 
requirement of the Pauli principle for the diquark correlation.
As for the vector coupling channel the diquark interaction channels can be 
obtained via Fierz transformation 
of the heavy gluon exchange model \cite{Buballa:2003qv}
which results in $G_D=0.75~G_S$ .
The scalar diquark interaction (\ref{diquark}) is attractive and can therefore
lead to diquark condensation at low temperatures and high densities,
according to the Cooper theorem \cite{Cooper:1956zz}.
As opposed to early works on diquark condensates and resulting color 
superconductivity which were based on perturbative one gluon exchange
\cite{Bailin:1983bm}, the renaissance of color superconductivity started
in 1997 was based on nonperturbative interaction models like the above NJL one
(see, e.g., \cite{Rapp:1997zu,Alford:1997zt,Blaschke:1998md}) 
and gave large diquark pairing gaps of the order of the fermion mass.
Therefore, the scalar diquark channel (\ref{diquark}) gives important 
contributions to the thermodynamics of cold, dense quark matter and cannot be
omitted in effective models of compact star matter.
These terms describe the standard NJL model as we use it for the present study.

However, several extensions are possible. We want to name 
a few of them which 
we consider important.
The first comes from the so called Kobayashi-Maskawa-'t Hooft (KMT)
interaction \cite{'tHooft:1973mm, Kobayashi:1970ji}
\begin{equation}
{\cal L}_K = -K[{\rm det}_f (\bar{q}(1+\gamma_5)q) 
+ {\rm det}_f(\bar{q}(1-\gamma_5)q)],
\end{equation}
where the determinants are taken in flavor space.
This determinant interaction is based on the single instanton solution and
gives access to the $U_A(1)$ anomaly
and the resulting $\eta - \eta^{'}$ mass splitting.
However, there  exist different
sources for this symmetry breaking  
\cite{Alkofer:2008et} which relate this aspect of low energy
QCD to nonperturbative field configurations in the gluon sector 
which (unlike the instantons) are also related to confinement
such as center vortices or a squeezed gluon condensate
\cite{Blaschke:1996dp,Pavel:1997mt}. 
To date, it is not clear which effects on the mean field thermodynamics 
of quark matter such alternative realizations of the  $U_A(1)$ anomaly
would have, if any. 
We consider any modeling of the  $U_A(1)$ symmetry breaking such as 
the KMT interaction (and its Fierz transformed interaction 
\cite{Steiner:2005jm,Hatsuda:2006ps,Abuki:2010jq} which involves the coupling
of chiral and diquark condensates) as part of an extension beyond the standard 
NJL-type model introduced above. 
Second, we mention the Polyakov loop potential $U(\Phi, \bar{\Phi})$ which is 
often added to NJL-type model Lagrangians in order to account for the 
existence of gluons and the phenomenon of (de)confinement.
In Polyakov gauge ($A_4 = \phi_3\lambda_3 + \phi_8\lambda_8$) the fields 
$\Phi$ and $\bar{\Phi}$ can be expressed as
\begin{equation}
 \Phi = \frac{1}{N_c} Tr_c \left\{\exp\left[i\int_0^\beta d\tau A_4(\tau)
\right]\right\}~.
\end{equation}
The form of the potential $U(\Phi, \bar{\Phi})$ in the presence of quarks, in 
particular at finite densities is not uniquely determined.
Several potential ans\"atze have been proposed in the literature 
\cite{Roessner:2006xn, Ratti:2005jh} 
where the temperature dependence of its coefficients has been determined by 
pure gauge lattice QCD.
A possible extension to finite chemical potentials has been proposed based on 
dimensionally consistent combinations of powers in terms of temperature and 
chemical potential \cite{Dexheimer:2009va, Blaschke:2010vj}.

A third extension  could be implemented by
a residual bag pressure which may even be
chemical potential dependent
to account for possible medium dependences 
of the gluon sector, as from a ``melting'' of the gluon condensate.
For recent models including a bag function along with a color superconducting 
NJL model of quark matter see, e.g.,  
\cite{Pagliara:2007ph,Bonanno:2011ch,Blaschke:2010vj,Lastowiecki:2012zz}.

Finally, we mention so-called crystalline (color) superconducting phases, 
also known as Larkin-Ovchinnikov-Fulde-Ferrell (LOFF) phases because of 
their similarity to condensed matter superconductors with magnetic 
impurities \cite{larkin:1964zz,Fulde:1964zz}, 
see Ref.~\cite{Anglani:2013gfu} for a recent review.
Initial investigations of LOFF phases \cite{Alford:2000ze,Bowers:2002xr} were performed with fixed quark masses and thus ignored the fact that a simultaneous selfconsistent solution of light and strange quark mass gap equations together with the pairing gap equations is essential for the phase structure itself.
This has been corrected later \cite{Ippolito:2007uz} and revealed that two-flavor color-superconducting LOFF phases can be energetically favored in compact stars and form stable hybrid star configurations with masses above 2 $M_\odot$ \cite{Anglani:2007aa,Ippolito:2007hn}.
The crystallinity of the 2SC phase may not essentially affect the question for the maximum mass of hybrid stars but can affect their cooling behavior and therefore our understanding of NS phenomenology
\cite{Sedrakian:2013pva}.

Let us now turn to the thermodynamical potential of the standard NJL model
for QM under NS constraints in the mean field approximation 
\cite{Blaschke:2005uj,Klahn:2006iw},

\small 
\begin{widetext}
\begin{equation}
 \Omega(T,\mu) = \frac{\phi_u^2 + \phi_d^2 + \phi_s^2}{8 G_S} 
- \frac{\omega_u^2 + \omega_d^2 + \omega_s^2}{8 G_V} 
+ \frac{\Delta_{ud}^2 + \Delta_{us}^2 + \Delta_{ds}^2}{4G_D}
- \int{\frac{d^3p}{(2\pi)^3}}\sum_{n=1}^{18}
  \left[E_n + 2T\ln\left(1 + e^{-E_n/T}\right)\right] 
+ \Omega_{l} - \Omega_0. 	 
\end{equation}
\end{widetext}
\normalsize
With $\Omega_l$  we added the lepton contributions (electrons and muons),
$\Omega_0$ guarantees zero pressure in the vacuum.
The extrema with respect to a variation of the  
meson and diquark mean fields $\phi_f, \omega_f$ and $\Delta_{fk}$ 
then define the gap equations
\begin{equation}
\frac{\partial \Omega}{\partial \phi_f}=
\frac{\partial \Omega}{\partial \omega_f}=
\frac{\partial \Omega}{\partial \Delta_{fk}}=0~,
\end{equation}
their solutions determine  thermodynamically stable equilibrium solutions.

\subsection{Model parameters}
As in-medium properties of quark matter are barely known,
an appropriate strategy to adjust free model parameters
is to describe well known vacuum properties of mesons (in general: hadrons).
In our model this is possible for the scalar coupling strength $G_S$
and the momentum cut-off $\Lambda$ which is necessary to regularize the 
divergent one-loop integrals. 
More parameters which enter
the model are the current quark masses $m_u$, $m_d$ and $m_s$.
For simplicity we set $m_u=m_d$.
In the chiral limit, for vanishing current quark masses, the pion and the 
kaon are true massless Goldstone bosons of the broken chiral symmetry.
The two parameters  $G_S$ and $\Lambda$ are then adjusted by the pion decay
constant $f_\pi=93$ MeV and, e.g., the chiral condensate 
$\langle \bar{u}u\rangle = - (240~{\rm MeV})^3$.  
The light and strange current quark masses are then fixed by  
the pion and kaon mass, respectively.
The details of this parameterization procedure
and a number of representative parameterizations
including the two sets used in the present work 
are found in \cite{Grigorian:2006qe}\footnote{Note, that
the strange current quark mass had to be corrected because of a mistake in the
kaon mass formula employed in \cite{Grigorian:2006qe}.
The corrected parameterization scheme has been implemented in an online tool 
developed by F.~Sandin and is available under
{\tt http://3fcs.pendicular.net/psolver}}.
The fitting procedure is not without ambiguities regarding the relation
between scalar coupling, light quark masses, remaining parameters
and the hadron properties the model is adjusted to.
In other words it is possible to describe this set of values
with different values of the scalar coupling.
Even though this does not give us arbitrary freedom to
choose a scalar coupling, slight variations are possible.

The values of the vector coupling strength $G_V$
and the diquark coupling strength $G_D$ 
can be constrained by the values they attain 
if all interaction channels originated from the 
Fierz rearrangement of a heavy gluon exchange model.
This defines their ratios to the scalar coupling strength
$\eta_V=G_V/G_S$ and $\eta_D=G_D/G_S$  
to be $\eta_V^{\rm Fierz}=0.5$
and  $\eta_D^{\rm Fierz}=0.75$, respectively.
These values actually result in a fair description of 
vector meson and nucleon masses (see, e.g., \cite{Ebert:1994mf}).
In this sense both values can be obtained from vacuum properties. 

We take a slightly different perspective, based on the idea
that both, vector meson and diquark interaction channels 
are particularly susceptible to density effects and 
thus become strongly renormalized under dense matter 
conditions which we apply for this model.
This situation is well known from the Walecka model for nuclear matter
where the scalar and vector meson couplings are adjusted to the 
phenomenological saturation properties of nuclear matter rather than
they reflect first principles.

Therefore, we use the above NJL model as an effective model
with $\eta_V$ and $\eta_D$ being free parameters.
A similar study we performed in \cite{Klahn:2011fb}
where we found hybrid EoS for a variety of $\eta_V$ and $\eta_D$  
which would predict NS with a QM core.

In this study we extend this scan
in order to explore the impact of the before mentioned
ambiguity regarding the precise choice of the scalar
coupling $G_S$.
For this purpose we chose two parameterizations from
Ref.~\cite{Grigorian:2006qe} which both describe
the same vacuum properties but differ by about 15\%
with respect to the scalar coupling.
The resulting parameterizations are shown in 
Tab. \ref{TAB:ModelParameters}.
Both sets have been used before,
Set A to obtain the results of Ref.~\cite{Sandin:2007zr},
Set B for example in Ref.~\cite{Klahn:2006iw}.
Note, that the constituent quark mass of Set B
is close to the nucleon mass divided by three and
therefore close to what we would consider as a
reasonable lower limit.

\begin{table}[h]
\begin{tabular}{|l|ccccc|}
 \hline
  &$G_S\Lambda^2$&$\Lambda\mbox{[MeV]}$& $m_{u}\mbox{[MeV]}$& $m_s\mbox{[MeV]}$& $M_u\mbox{[MeV]}$\\ 
\hline\hline
 Set A  &  $2.319$ & $602.30$ & $5.500$ & $112.00$ & 367.5 \\
 Set B  & $2.176$ & $629.54$ & $5.277$ & $135.88$ & 330.0 \\ \hline 
\end{tabular}
 \caption{NJL model parameterization used in the present study, taken from 
Ref.~\cite{Grigorian:2006qe}.
 $M_u=m_u+\phi_u$ is the constituent light quark mass in vacuum.}
 \label{TAB:ModelParameters}
\end{table}

\subsection{Quark-hadron phase transition}
The NJL model as we use it describes the thermodynamics of deconfined quark 
fields and does not account for the formation of hadrons due to confinement.
A simple and feasible way to 
describe the transition
from nuclear to quark matter is to take advantage of an 
independently calculated equation of state of nuclear matter.
With this input the phase transition between the
two phases is constructed based on the standard Gibbs rules of phase equilibrium.
The nature of the QCD phase transition is not clear,
there are ongoing discussions whether it is strongly first order
(a Maxwell-like transition), permits the formation of regions
with mixed phases of hadrons and quarks and even pasta-like phases
or even a cross over transition with no critical end point
in the QCD phase diagram at all.
Here, we assume a strong first order phase transition,
modelled by a Maxwell-construction. 
For details and the specifics of this construction under the constraints of 
$\beta-$equilibrium as well as electric and color charge neutrality in the 
presence of diquark condensates see, e.g., 
Refs.~\cite{Baldo:2002ju,Grigorian:2003vi,Shovkovy:2003ce}.

For our analysis 
we apply the nuclear Dirac-Brueckner Hartree-Fock (DBHF) EoS 
which has proven to perform reasonably well
for describing nuclear matter saturation properties and kaon data  
\cite{Fuchs:2003zn}
as well as NS properties \cite{Klahn:2006ir}
even  though it tends to behave too stiff
above densities of about 3.5 times saturation density.
On the other side, this stiffness occurs in
a region where QM degrees of freedom are not
unlikely to be the only ones which are relevant.
Amongst other reasons we prefer the DBHF EoS 
because it is based on a relativistic and microscopical 
description of many-particle interactions.
It starts from a given free
nucleon-nucleon interaction (the relativistic Bonn A potential) 
fitted to nucleon-nucleon scattering data and deuteron properties. 
In ab initio calculations based on many-body techniques one then
derives the nuclear energy functional from first principles,
i.e., treating short-range and many-body correlations explicitly.
In the relativistic DBHF approach
the nucleon inside the medium is dressed by the
self-energy based on a T-matrix. The in-medium T-matrix
as obtained from the Bethe-Salpeter equation
plays the role of an effective two-body interaction
which contains all short-range and many-body correlations
in the ladder approximation.
As we have shown in the context of hybrid EoS the rather stiff behavior at 
high densities is not necessarily relevant if the phase transition to QM 
occurs at low enough densities of about three to four times saturation 
density \cite{Klahn:2006iw}.

\section{Results}

For both sets, A and B, with different parameterizations
regarding the scalar coupling strengths $G_S$ 
we calculated the full QM EoS for eight values of the
effective vector coupling $\eta_V$ between $0$ and $0.7$ in
steps of $0.1$. For the effective diquark coupling we
chose a stepwidth of $0.02$ in the interval $[0.8,0.94]$
and $0.01$ in the interval $[0.94,1.15]$ equal to
29 different values.
These choices provided a sufficient coverage of the range of parameters
which can result in stable hybrid star configurations.
Additionally we performed the same amount of calculations
for symmetric matter.
This gives a total of $928$ different QM EoS we computed,
where each required about two hours of computing time on
a 2.7 GHz quad-core Opteron(tm) processor of which we had ten available.

With the resulting hybrid EoS we solved the TOV equations 
and obtained mass-radius (M-R) and mass-central density relations (M-n) 
relations for spherically symmetric compact stars.
The computational effort for this part has been negligible
despite the very large number of calculated neutron stars.

In the following we explore the relation between the
free parameters 
($\eta_D$, $\eta_V$, and the scalar coupling
strength $G_S$) of the introduced NJL model EoS
and the resulting NS characteristics
(maximum NS masses, critical NS masses for
the phase transition, and the corresponding M-R relations).
We will highlight the parameter regions where agreement with
the observations of $2M_\odot$ pulsars and flow measurements in HIC
could been obtained.

\subsection{Analysis of the $2 {\rm M}_\odot$ constraint}

\begin{figure}[h]
 \includegraphics[scale=0.33,angle=-90]{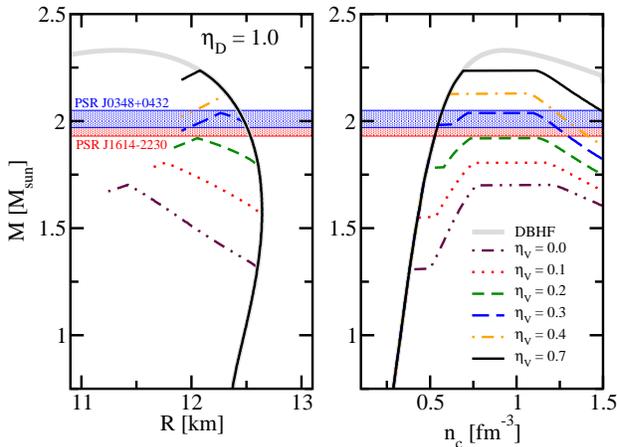}
 \caption{(Color online.) 
Mass-radius and mass-central density sequences for varying vector 
coupling strength $\eta_V$ 
at fixed $\eta_D=1.0$ 
for set A.}
 \label{FIG:EtaDScan}
\end{figure}

\begin{figure}[h]
 \includegraphics[scale=0.32,angle=-90]{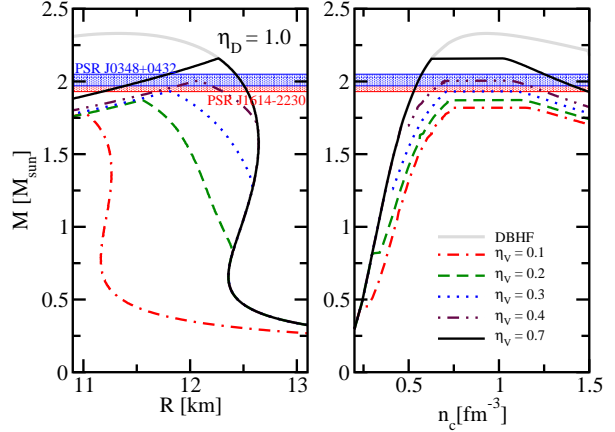}
 \caption{(Color online.) Same as Fig.~\ref{FIG:EtaDScan} for set B.}
 \label{FIG:EtaDScan2}
\end{figure}

In order to illustrate the general influence of the coupling constants
$\eta_D$ and $\eta_V$ we will first vary only one of them and keep the other constant.
We start with the diquark coupling at a constant value of $\eta_D=1.0$ 
and vary the vector coupling $\eta_V$. 
As it is expected from an repulsive interaction channel
increasing $\eta_V$ increases the stiffness of the QM EoS.
As a consequence a higher value of $\eta_V$ leads to higher  densities for the onset of the phase transition and simultaneously to higher maximum NS masses.
The same general behavior holds for both sets, A and B,
as we illustrate in Figs.~\ref{FIG:EtaDScan} and \ref{FIG:EtaDScan2}.
Note, that despite this general stiffening of the EoS
with increasing vector coupling, the transition 
densities for the same sets of  values ($\eta_D,\eta_V$) differ 
between both sets.
This difference is most pronounced at small values of $\eta_D$
where Set A shows an onset of QM at higher densities than Set B.

\begin{figure}[h]
 \includegraphics[scale=0.33,angle=-90]{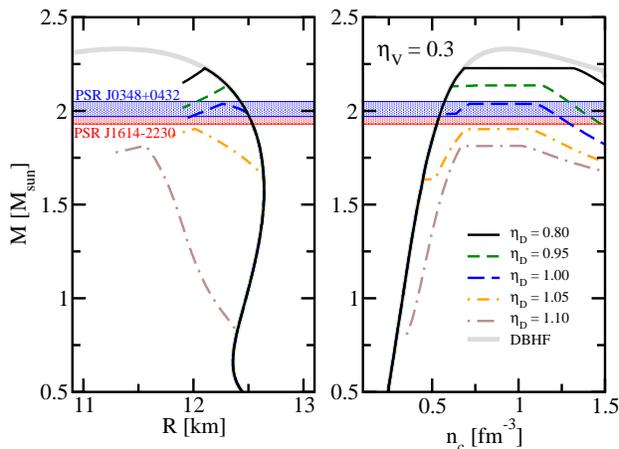}
 \caption{(Color online.) 
Mass-radius and mass-central density sequences for varying diquark coupling 
strength $\eta_D$ at fixed $\eta_V=0.3$ 
for set A.}
 \label{FIG:EtaVScan}
\end{figure}

\begin{figure}[h]
 \includegraphics[scale=0.32,angle=-90]{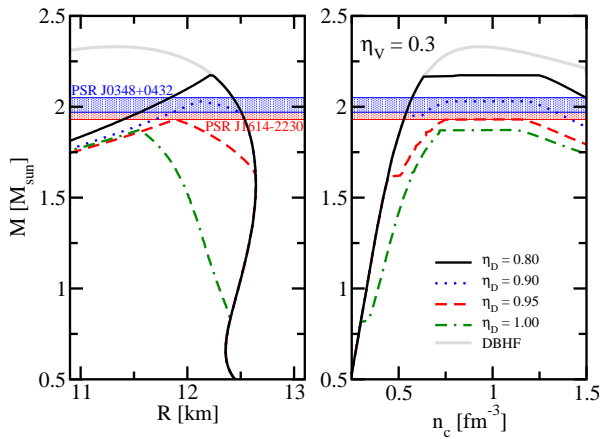}
 \caption{(Color online.) Same as Fig.~\ref{FIG:EtaVScan} for set B.}
 \label{FIG:EtaVScanB}
\end{figure}

Next, we perform a variation of $\eta_D$
at a fixed value of $\eta_V=0.3$.
Again set A and B have the same systematic behavior with only quantitative differences, as shown in the Figs. \ref{FIG:EtaVScan} and \ref{FIG:EtaVScanB}. 
Increasing  the coupling strength in the diquark interaction channel 
lowers the critical density  and at the same time the maximum attainable mass 
is lowered due to 
the resulting
softening of the EoS.


Comparing set A and B at the same values of parameters $\eta_D$ and $\eta_V$
shows that a lower constituent quark mass (set B) results in an earlier 
transition to QM.
We point out that because both, $\eta_D$ and $\eta_V$, are defined as the 
ratio of the corresponding coupling strengths to the scalar coupling strength 
$G_S$ we explicitly do not compare equal coupling strengths in the coupling 
channels.

\begin{figure}[h]
 \includegraphics[scale=0.33,angle=-90]{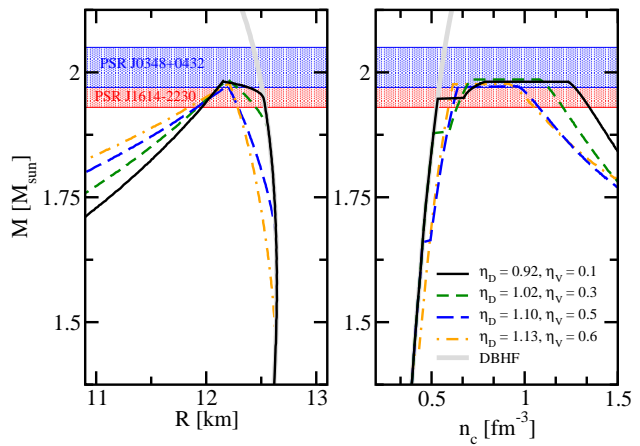}
 \caption{(Color online.) 
Systematics of mass-radius and mass-central density curves for 
set A with the parameter pairs ($\eta_D,\eta_V$)
chosen such that
the maximum mass equals $1.97~M_\odot$, the mass of PSR J1614-2230.
}
 \label{FIG:ConstrScan}
\end{figure}

\begin{figure}[h]
 \includegraphics[scale=0.32,angle=-90]{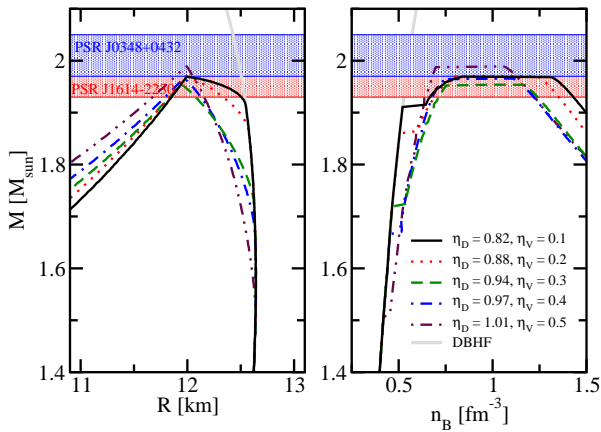}
 \caption{(Color online.) Same as Fig.~\ref{FIG:ConstrScan} for set B.}
 \label{FIG:ConstrScan2}
\end{figure}

In Figs.~\ref{FIG:ConstrScan} and \ref{FIG:ConstrScan2} we present M-R curves 
for parameter sets chosen such that the maximum obtained mass is 
$1.97~M_\odot$, corresponding to the mass expectation value of 
PSR J1614-2230 \cite{Demorest:2010bx}
and to the lower limit of the $1\sigma$ band of the mass measurement for 
PSR J0348+0432 \cite{Antoniadis:2013pzd}.

One clearly observes the general tendency that 
when increasing the vector coupling one would have to increase also the 
diquark coupling in order 
to keep the maximum NS mass at a constant value.
This represents the earlier stated fact, that $\eta_V$ stiffens and $\eta_D$ 
softens the EoS.
Therefore, we find 
that the constraint of a given NS mass is fulfilled for monotonously rising 
functions in the $\eta_D - \eta_V$ parameter plane.

We note, that for any hybrid EoS which reproduces the maximum mass 
of PSR J1614-2230 the radius of the most massive configuration does not 
depend significantly on the chosen parameters and is around 12 km.

\begin{figure}[h]
 \includegraphics[scale=0.32,angle=-90]{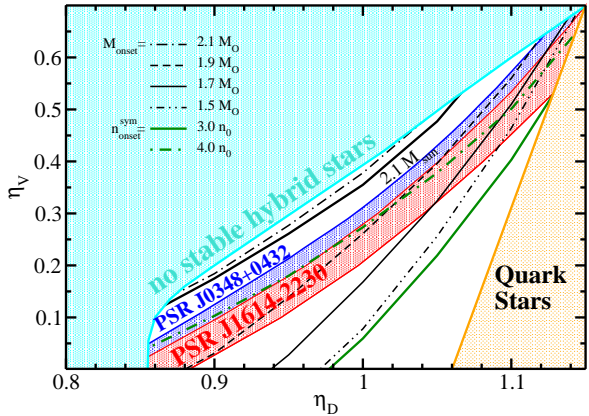}
 \caption{(Color online.) 
Full analysis of 
hybrid NS with QM core in
the $\eta_V$-$\eta_D$ parameter space 
for set A.
The colored hatched regions denote parameter pairs where no stable hybrid
 stars are possible (cyan), the entire star is composed of quark matter
 (orange) and where the maximum mass for the hybrid EoS is contained in the
 $1\sigma$ band of the mass measurement for PSR J1614-2230 (red) or 
PSR J0438+0432 (blue).
The bold green lines denote given densities for the QM onset in symmetric 
matter: $3~n_0$ (solid) and $4~n_0$ (dash-dotted).
The thin black lines stand for given NS masses at QM onset: $2.1~M_\odot$ 
(dash-dotted),  $1.9~M_\odot$ 
(dashed),  $1.7~M_\odot$ 
(solid), and  $1.5~M_\odot$ (dash-double dotted).  
}
 \label{FIG:MonsterPlot}
\end{figure}

\begin{figure}[h]
 \includegraphics[scale=0.32,angle=-90]{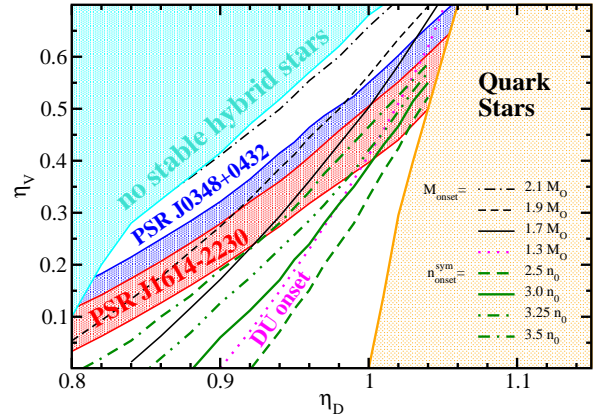}
 \caption{(Color online.) Same as Fig.~\ref{FIG:MonsterPlot} for Set B;
additional densities for the QM onset in symmetric matter (see legend) 
indicate the onset of the direct Urca (DU) cooling process in NS matter
at $1.3~M_\odot$ according to the hadronic DBHF EoS (bold dotted magenta 
line).
}
 \label{FIG:MonsterPlot2}
\end{figure}

The discussed results are summarized in Fig.~\ref{FIG:MonsterPlot} and 
Fig.~\ref{FIG:MonsterPlot2} for Set A and Set B, respectively.
The red band covers all possible parameterizations
which result in a maximum mass equal to the mass of PSR J1614-2230
while the blue band corresponds to the mass of PSR J0438+0432.
Non-solid black lines refer to a certain NS mass (given in the legend) 
at which QM appears in the NS core.
This information is useful to estimate the amount of QM in a given
parameterization.
Note, that if one follows the red (or blue)
band from the left to the right
the corresponding critical NS mass, where QM appears first,
decreases. This just illustrates that one can expect larger QM
cores in NS configurations with higher values of $\eta_D$ (and consequently 
$\eta_V$) if the different EoS all result in the same maximum NS mass.
Of course, the reason for this is a lowering of the critical density
along the 
red (blue) 
band from the left to the right.
The solid black line, labelled with $2.1~M_\odot$
denotes the border between model 
parameter regions for which the maximum NS mass is below 
or beyond $2.1~M_\odot$. 

At any given $\eta_D$, heavier 
configurations are found at higher values for $\eta_V$.
For parameterizations within the cyan region we do not obtain any stable 
hybrid NS configuration.
Even though all these solutions are purely hadronic, the maximum masses
in this region can differ, as 
the phase transition to quark matter can occur at central densities
below the central density corresponding to the maximum mass of the
purely hadronic EoS. In this case a transition to QM only results
in a lowering of the maximum mass of purely hadronic NS due to 
the instability of the hybrid star configurations.
The light 
orange
region denoted as ''Quark Stars'' deserves a separate discussion. 
In this domain we find quark matter favored over hadronic matter
at all densities (or chemical potentials).
Therefore, a phase transition does not occur.
Even though we consider the total missing of a phase transition, in particular 
at low densities, as an artefact of our QM model one
can conclude, that in this domain the phase transition is taking
place at extremely low densities. Therefore, the label 'quark stars'
is justified even though the distinction between hybrid configurations
and pure quark stars is not as strict as it appears.
Near this border we observe a phenomenon known as masquerade effect \cite{Alford:2004pf}
where the EoS for quarks and hadrons are nearly identical 
and one therefore obtains almost identical  M-R curves (hence the “masquerade”) for the purely hadronic and the hybrid EoS. 
In this scenario one can sometimes observe multiple crossings of the
hadronic and QM EoS in the P-$\mu$ plane. 
This can be interpreted
as an indicator for a crossover transition from one phase to the other,
characterized by identical EoS in the transition region.
Certainly, a Maxwell construction as we have performed 
is not suited to address this scenario.

The green lines in Figs.~\ref{FIG:MonsterPlot} and \ref{FIG:MonsterPlot2} 
refer to the critical density of symmetric 
hadronic matter where the phase transition to QM takes place (the values
are given in the plot).
As this concerns properties of symmetric matter we will discuss this
in more detail in the following section.

\subsection{Connection to symmetric matter and the flow constraint}
While matter in a NS due to the established $\beta$-equilibrium and
electric charge neutrality is in general highly isospin asymmetric,
in particular in the 
NS core, the 
matter in HIC is fairly isospin symmetric.
Understanding the properties of symmetric matter under the HIC conditions of 
high temperature and density is of great importance 
for the exploration and the understanding of a wide range of phenomena.
Focussing on the QCD phase transition one can divide measured
observables roughly into those which are are highly sensitive to a phase 
transition and those which are not. 
Most observables concerning particle yields fall into the latter category. 
The most promising for the exploration of thermodynamic properties of matter 
in heavy ion collisions are processes associated with the hydrodynamic 
expansion of the fireball and connected to anisotropies in the observed 
particle distributions. 
An 
interpretation of the measurements in terms of thermodynamic properties 
is a difficult and highly involved task that suffers from systematic 
uncertainties. 
Attempts have been made to analyze elliptic flow data
in order to specify a region in the pressure-density plane 
which provides upper and lower limits for the pressure at
a given density \cite{Danielewicz:2002pu}. 

\begin{figure}[h]
 \includegraphics[scale=0.32,angle=-90]{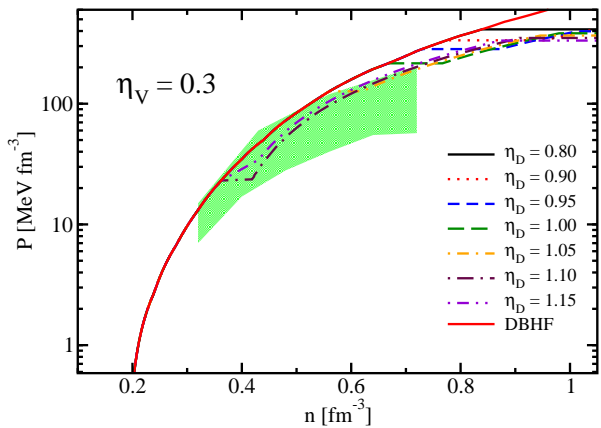}
 \caption{(Color online.) 
Symmetric hybrid EoS at different $\eta_D$ for fixed $\eta_V=0.3$ 
in comparison to the flow constraint for Set A.}
 \label{FIG:FlowVScan}
\end{figure}

\begin{figure}[h]
 \includegraphics[scale=0.32,angle=-90]{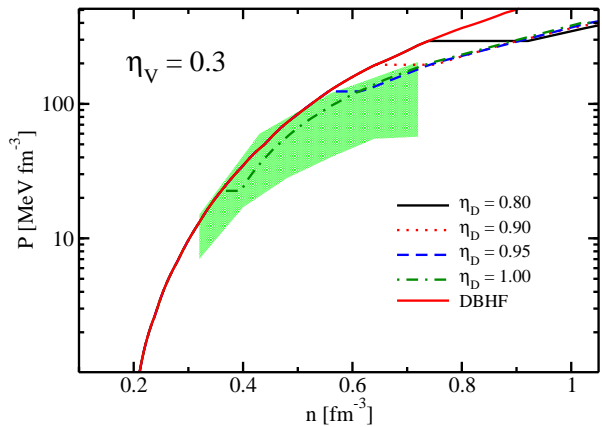}
 \caption{(Color online.) Same as Fig. \ref{FIG:FlowVScan} for Set B.}
 \label{FIG:FlowVScan2}
\end{figure}

Our conclusions for the qualitative dependence of the deconfinement phase 
transition on the $\eta$-parameters of our model hold in the same way for 
symmetric matter as discussed in the previous section for NS matter.
We remind, that a higher value of $\eta_V$ leads to an onset of quark matter 
at higher densities and conversely a higher value of $\eta_D$ leads to an 
earlier transition. 
An increase of the constituent quark mass 
(by a larger scalar coupling strength $G_S$)
leads to lower transition densities. 
Figs.~\ref{FIG:FlowVScan} and \ref{FIG:FlowVScan2} 
(for set A and B, respectively) show the EoS for symmetric matter at a fixed 
value of $\eta_V=0.3$ with different coupling strengths in the diquark channel.
The results are plotted on top of the flow constrained region (green area).
It is clearly visible that the red solid plotted hadronic DBHF EoS would
violate the flow constraint at about $n=0.55~{\rm fm}^{-3}$,
corresponding to 3.5 times the saturation density $n_0$.
While describing the phase transition by performing a Maxwell construction
the occurrence of a quark branch in the hybrid EoS necessarily leads to 
a softening of the EoS 
in the corresponding domain.
In our scenario this turns out to be of advantage, as 
the violation of the flow constraint by DBHF at high densities is 
corrected for the hybrid EoS, given the phase transition
occurs around the density $n\simeq 0.55~{\rm fm}^{-3}$.
In this particular example with $\eta_V=0.3$ this holds for 
$\eta_D \gsim 1.05$
(Set A, Fig.~\ref{FIG:FlowVScan}) or 
$\eta_D \gsim 0.95$
(Set B, Fig.~\ref{FIG:FlowVScan2}), respectively.

While the value of $\eta_V$ in the previous paragraph has been chosen 
arbitrarily one 
could ask for a justification of this parameter choice.
As our aim is to find connections between NS and HIC observables
we discuss the question what consequences arise for the phase
transition in HIC if one would know that QM exists at least in 
the heavier compact stars. 
Because the NS mass increases with the central density the opposite 
scenario - QM exists {\it only} in less massive NS - is not realistic for 
regular NS. 
This might be different for '3rd family' stars which arise from twin solutions 
where one can find a second stable branch of exotic NS with smaller radii
and masses below those we discuss.

\begin{figure}[h]
 \includegraphics[scale=0.32,angle=-90]{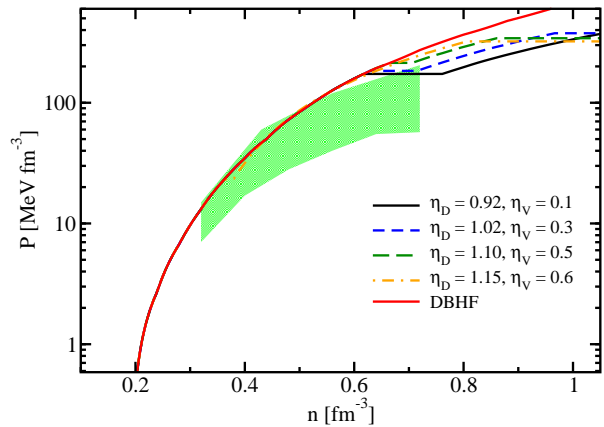}
 \caption{(Color online.) Symmetric hybrid EoS
for different pairs of ($\eta_V$, $\eta_D$), all 
describing a maximum NS mass equal to that of PSR J1614-2230. (Set A.)}
 \label{FIG:FlowConstrScan}
\end{figure}

\begin{figure}[h]
 \includegraphics[scale=0.32,angle=-90]{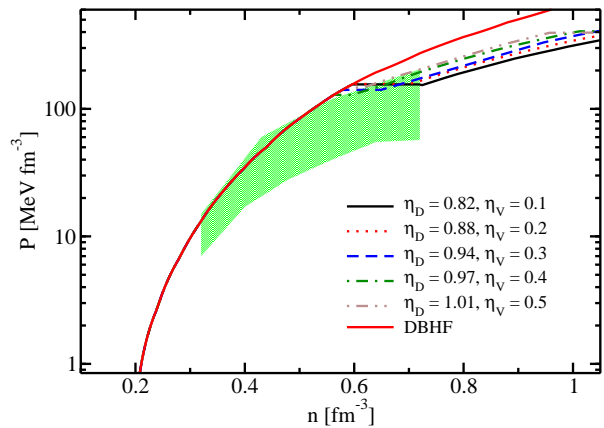}
 \caption{(Color online.) Same as Fig. \ref{FIG:FlowConstrScan} for Set B.}
 \label{FIG:FlowConstrScan2}
\end{figure}

In Fig.~\ref{FIG:FlowConstrScan} and Fig.~\ref{FIG:FlowConstrScan2} 
(representing Set A and B, respectively) 
we show symmetric EoS with parameters which under NS constraints
result in a maximum masses of $1.97~M_\odot$, the mass of PSR J1614-2230.
In both cases we find similar values for the critical density,
all in the vicinity of the density where 
the DBHF EoS would begin
to violate the flow constraint.
For Set B, the parameterization with a smaller value of the constituent quark 
mass, the phase transition is shifted towards slightly lower densities. 
As this brings Set B into better agreement with the flow constraint, this 
slight change is meaningful.
In these Figures the phase transition in symmetric matter appears 
to take place at an almost constant value.
As the EoS parameterizations with respect to $\eta_V$ and $\eta_D$ are
very different and just agree in reproducing the same maximum mass 
(chosen to be 
$1.97~M_\odot$, compatible with the measured masses of both $2~M_\odot$ 
pulsars, PSR J1614-2230 \cite{Demorest:2010bx} and PSR J0438+0432 
\cite{Antoniadis:2013pzd}, within the $1\sigma$ range) 
this gives the following interesting result. 

If it turns out, that the mass of PSR J1614-2230 is close
to the maximum mass a NS can maintain, then the actual parameters of
the QM model are of minor relevance for the approximate value of
the critical density in symmetric matter.
This holds under the assumption that QM does exist in the core of this 
pulsar.
If we assume the opposite, claiming that QM does not exist in any NS,
our observation is still useful as the almost parameter
independent critical density we found can be interpreted as the absolute
lower limit to the actual critical density in symmetric matter. 
The phase transition in HIC under no circumstances
can occur at lower densities as this would result in NS with quark matter core.

However, more valuable from the perspective of how to constrain our QM model 
is the first situation, where at least some NS are hybrid stars.
If confirmed, this would greatly constrain the possible parameter range of our 
model.
To illustrate this we refer to Fig.~\ref{FIG:MonsterPlot} and 
Fig.~\ref{FIG:MonsterPlot2}.
In both cases the allowed region of parameters is limited
from below
by the lower edge of the red band and
from above
by the lower edge of the region with no stable hybrid solutions.
If the existence of QM in compact stars is excluded
by whatever reason a realistic QM EoS can, of course, be found only
in the cyan hatched region of parameterizations which result in unstable
hybrid configurations and configurations with a quark matter onset beyond
the central density for the maximum mass of the purely hadronic EoS.
Looking closely, one realizes that this argumentation works best for Set A.
In the corresponding Fig.~\ref{FIG:MonsterPlot} the 
red (blue) band,
denoting configurations with maximum masses 
in the $1\sigma$ band of mass measurements for the $2~M_\odot$ pulsar
PSR J1614-2230 (PSR J0438+432)
runs almost parallel to the dash-dotted green curve which represents
parameterizations with a constant critical density of $n_c=4.0~{\rm n}_0$.
For Set B in Fig.~\ref{FIG:MonsterPlot2} a similar discussion could hold for
smaller values of $\eta_D$ (and therefore $\eta_V$). 
However, for values close to $\eta_D=1.0$ one can find very different
critical densities in symmetric matter for slightly different 
parameterizations.

We stated early in this paper, that one of the big unknowns
in our analysis is the hadronic EoS. 
In this sense our analysis 
is not complete and should be repeated
with a wide range of different nuclear EoS in order to get a better idea about 
the possible 
spectrum of hybrid EoS obtained within the class of two-phase models with 
basic superconducting NJL quark matter.
For now, we can just pretend, that DBHF is a fairly
realistic hadronic EoS. A last, and very interesting result has
to be understood from this perspective. If we strictly require
that our QM EoS {\it has to} repair the violation of the flow constraint by
DBHF this means that the phase transition in symmetric matter has to take
place below a density of about 
$3.25~n_0$.
In the case of Set A as visible in Fig.~\ref{FIG:MonsterPlot} this is 
difficult to achieve but not impossible. 
There is a small region at $\eta_D\approx 1.1$, $\eta_V\approx 0.55$ which would permit this. 
In this region of the parameter space, 
the size of the QM core in NS is large, 
the transition density consequently very low.
{\it If} DBHF is a realistic EoS for low densities in NS matter and
up to 
$3.25~n_0$
in symmetric matter and {\rm if} the upper bound of the flow constraint is 
reliable the existence of heavy NS with QM content
provides a very tight constraint on both, $\eta_D$ and $\eta_V$.
The situation is a bit less extreme for Set B, as seen in 
Fig.~\ref{FIG:MonsterPlot}. In this case one would expect $\eta_D$
to be larger than 0.975 but not larger than 1.05.
The vector coupling $\eta_V$ in this domain can vary widely
between about 0.4 and 0.6.

We point out again, that Set A corresponds to higher constituent quark masses than Set B. Constituent quark masses below these for Set B
are barely realistic. In this sense we find, that high constituent quark masses are less easily brought into agreement with DBHF as an EoS which
does not fulfil the flow constraint. Ignoring this it seems easier
for the same set to derive lower limits for the onset of quark matter
in HIC.

\section{Conclusions}
The major conclusion of this investigation is that 
{\it\bf 
the existence of quark matter in neutron stars is not excluded by any current 
data}.
We described a wide number of parameterizations which result 
in NS configurations with a quark matter core 
that reach 
masses up to $2.1~M_\odot$ and thus fulfill the constraint from the new 
mass measurements of PSR J1614-2230 and PSR J0438+432
Some of our parameterizations resulted in NS with quark cores containing 
half of the total mass of the star.
The biggest unknown in our investigation is the hadronic EoS
which determines the maximum NS mass and the precise value of
any quantity we investigated.

For one of the investigated sets (Set A) we found 
a strong correlation between the onset of QM in symmetric
matter and the maximum mass of the hybrid star sequence.
This particular model set up implies, that the critical density
in symmetric matter is larger than or equal to 
four times saturation density.
With our choice of the hadronic EoS (DBHF) this implies
a violation of the flow constraint 
unless we choose the above mentioned set of very strong couplings
$(\eta_D,\eta_V)=(1.1,0.55)$ .
If there is quark matter in neutron stars the transition density
cannot be much larger because otherwise hybrid NS configurations turn
unstable. On the other hand, even if there is no QM in NS the onset
density cannot be lower, as this would theoretically result in stable 
hybrid configurations.

If heavy-ion collision experiments provide precise information about the
transition density in symmetric matter, one can favor or disfavor the existence of quark matter in NS cores.
This correlation is less distinct for Set B with different scalar coupling
and a very small resulting constituent quark mass.
In this case we still estimate the lowest transition density
in symmetric matter to have a value of about $2.5~n_0$.
However, a similar precise statement about the largest critical density
which could still be in agreement with the hypothetical existence of
QM cores in NS is out of scope considering the uncertainties
due to the badly known hadronic EoS.
Assuming, that DBHF is 'the' EoS and that a transition to QM repairs
the violation of the flow constraint the upper limit is 
the density where this violation begins, namely $3.25~n_0$.

To bring NS phenomenology and HIC flow data in agreement high values of the diquark and scalar coupling are required.
In particular the vector interaction channel 
should have a coupling value of  $\eta_V\approx0.5$ or above.
For the diquark channel values of $\eta_D$ around $1.0$ result
in hybrid stars configurations which are in good agreement with 
our data.

A lower  constituent quark mass leads to an earlier onset of quark matter.
In this context such parameterizations favor the existence of QM in NS.
The fulfillment of the flow constraint is easier to obtain with a smaller constituent quark mass.
However, the outcome of this analysis depends strongly on the hadronic EoS.
Ignoring the violation of the flow constraint which is induced
to this analysis by the DBHF EoS allows a  wide range of $\eta_D$ and $\eta_V$ for both, Set A and Set B. 
Both coupling parameters, $\eta_V$ and $\eta_D$, are correlated due to 
observed high mass NS.

Summarizing our conclusions, we find that the observation of the 
$2~M_\odot$ pulsars PSR J1614-2230 and PSR J0438+0432
provides strong constraints for the occurrence of quark matter in compact 
stars.
Within chiral quark models of the NJL type the occurrence of QM in NS requires 
that QM must be color superconducting with a rather large diquark coupling
constant and requires repulsive forces coming from a non vanishing mean field 
in the vector meson interaction channel.

These conclusions are drawn from a mean-field description of quark 
matter within a two-phase description of quark-hadron hybrid matter. 
This analysis is not based on first principles 
but on the other hand represents what is the current state of the art
concerning the modelling of matter at high densities and low temperature.

\section*{Acknowledgements}

This work was supported in part by ``CompStar'' a research networking programme
of the European Science Foundation and by a grant from the Polish Ministry for 
Science and Higher Education (MNiSW) supporting it. T.K. and D.B. acknowledge 
partial support by Narodowe Centrum Nauki (NCN) within the ``Maestro'' 
programme under contract No. DEC-2011/02/A/ST2/00306. 
Further support came from the Russian Fund for Basic Research under
grant number 11-02-01538-a (D.B.), the EU framework programme FP7 
``hadronphysics3'' (T.K.), a research grant number 2291/M/IFT/12 from 
the Faculty of Physics and Astronomy at the University of Wroclaw and by 
European Social Fund under project number POKL.04.01.01-00-054/10-00(R.{\L}.).

\end{document}